\documentclass{edm_article_modified}
\usepackage{balance} 
\usepackage[utf8]{inputenc} 
\usepackage{subcaption}
\usepackage{multirow}
\usepackage{booktabs}
\usepackage{colortbl}
\usepackage{diagbox}
\usepackage{xspace}
\usepackage{enumitem} 
\usepackage{mdframed} 
\usepackage{pifont} 
\usepackage[numbers,sort&compress]{natbib}
\usepackage{url} 
\usepackage{xcolor}
\usepackage{enumitem}
\usepackage{tikz} 
\newcommand{\notick}{\ding{55}}
\newcommand{\yestick}{\ding{51}}
\definecolor{urlcolor}{rgb}{0.0, 0.53, 0.74}

\newcommand{\textcode}[1]{{\fontfamily{cmtt}\selectfont #1}\xspace}

\newcommand{\DSLRepeat}{\textcode{\textsc{Repeat}}}
\newcommand{\DSLRepeatUntil}{\textcode{\textsc{RepeatUntil}}}

\newcommand{\DSLIf}{\textcode{\textsc{If}}}
\newcommand{\DSLIfElse}{\textcode{\textsc{IfElse}}}

\newcommand{\ourtest}{\textsc{ACE}}
\newcommand{\ourtestheader}{ACE}
\newcommand{\ourtestone}{\textsc{ACE[01-07]}}
\newcommand{\ourtesttwo}{\textsc{ACE[08-14]}}
\newcommand{\ourtestthree}{\textsc{ACE[15-21]}}
\newcommand{\hocmaze}{\textsc{HoCMaze}}
\newcommand{\apply}{\textsc{Applying}}
\newcommand{\analyze}{\textsc{Analyzing}}
\newcommand{\evaluate}{\textsc{Evaluating}}
\newcommand{\create}{\textsc{Creating}}

\begin{document}

\title{Analyzing--Evaluating--Creating: Assessing Computational Thinking and Problem Solving in Visual Programming Domains\thanks{This extended version of the SIGCSE 2024 paper includes all $21$ test items from \ourtest{} along with their answers in the appendix.}}





\numberofauthors{3}
\author{
\alignauthor
		Ahana Ghosh\\
       \affaddr{MPI-SWS}\\
       \affaddr{Saarbr{\"u}cken, Germany}\\
       \email{gahana@mpi-sws.org}
\alignauthor
		Liina Malva\\
       \affaddr{MPI-SWS}\\
       \affaddr{Saarbr{\"u}cken, Germany}\\
       \affaddr{Tallinn University}\\
       \affaddr{Tallinn, Estonia}\\
       \email{liina.malva@tlu.ee}
\alignauthor
		Adish Singla\\
       \affaddr{MPI-SWS}\\
       \affaddr{Saarbr{\"u}cken, Germany}\\
       \email{adishs@mpi-sws.org}
}



\maketitle


\begin{abstract}
Computational thinking (CT) and problem-solving skills are increasingly integrated into K-8 school curricula worldwide. Consequently, there is a growing need to develop reliable assessments for measuring students' proficiency in these skills. Recent works have proposed tests for assessing these skills across various CT concepts and practices, in particular, based on multi-choice items enabling psychometric validation and usage in large-scale studies. Despite their practical relevance, these tests are limited in how they measure students' computational creativity, a crucial ability when applying CT and problem solving in real-world settings. In our work, we have developed \ourtest{}, a novel test focusing on the three higher cognitive levels in Bloom's Taxonomy, i.e., \analyze{}, \evaluate{}, and \create{}. \ourtest{} comprises a diverse set of $7$\ensuremath{\times}$3$ multi-choice items spanning these three levels, grounded in elementary block-based visual programming. We evaluate the psychometric properties of \ourtest{} through a study conducted with $371$ students in grades $3$--$7$ from $10$ schools. Based on several psychometric analysis frameworks, our results confirm the reliability and validity of \ourtest{}. Moreover, our study shows that students' performance on \ourtest{} positively correlates with their performance in solving tasks on the \emph{Hour of Code: Maze Challenge} by Code.org.
\end{abstract}

\section*{Keywords}
Computational Thinking, Assessment Tools, Bloom's Taxonomy, Visual Programming


\section{Introduction}\label{sec.intro}

Computational thinking (CT) is emerging as a critical skill in today's digital world. According to the work of~\citep{wing2006computational}, ``computational thinking involves solving problems, designing systems, and understanding human behavior, by drawing on the concepts fundamental to computer science''. Several works have also discussed the multi-faceted nature of CT and its broader role in the acquisition of creative problem-solving skills~\citep{DBLP:journals/jeric/GroverBBEDS17, chevalier2020fostering}. As a result, CT is being increasingly integrated into K-$8$ curricula worldwide~\citep{grover2013computational, DBLP:conf/chi/LoksaKJOMB16}. With the growing integration of CT at all academic stages, there has also been a surge in demand for validated and reliable tools to assess CT skills, especially at the K-$8$ stages~\citep{DBLP:journals/ijcses/LockwoodM18, DBLP:journals/corr/abs-2203-05980}. These assessment tools are essential for tracking the progress of students, guiding the design of curricula, and supporting teachers as well as researchers to assist students in the 
acquisition of CT skills~\citep{DBLP:conf/educon/Zapata-CaceresM20, chevalier2020fostering, gonzalez2015computational, DBLP:journals/ijcses/LockwoodM18}. 

Prior work has proposed several assessments that measure students' CT during their K-$8$ academic journey. On the one end, several portfolio-based assessments have been proposed that measure students' CT through projects in specific programming environments~\citep{DBLP:journals/ce/TangYLHZ20}. Although portfolio-based tests provide \emph{open-ended projects} to capture students' analytical, evaluative, and creative skills, they are challenging to implement and interpret on a larger scale~\citep{roman2019combining, DBLP:journals/corr/abs-2203-05980}. On the other end, several diagnostic assessment tools have been proposed that measure CT in the form of \emph{multiple-choice items}~\citep{DBLP:journals/chb/Roman-GonzalezP17, DBLP:journals/corr/abs-2203-05980, DBLP:journals/jeric/Lai22, DBLP:conf/sigcse/BockmonB23}. These assessment tools are preferred for their practicality in large-scale administration and suitability for both pretest and posttest conditions~\citep{roman2019combining}. However, scalability comes at the cost of limiting the ability to effectively measure students' computational creativity. Thus, there is a need to develop multi-choice tests that also capture students' computational creativity.

\begin{figure*}
\captionsetup[subfigure]{aboveskip=3pt,belowskip=3.25pt}
    \begin{subfigure}{0.49\textwidth}
    \scalebox{0.68}{
        \centering
        \setlength\tabcolsep{1.2pt}
        \renewcommand{\arraystretch}{1.51}  
        \begin{tabular}{l || cc || cc || ccc}
        \toprule 
        \multirow{3}{*}{\diagbox[width=10em]{\textbf{Concept}}{\textbf{ Level\ \ }}}
        & \multicolumn{2}{c||}{\textbf{{\apply{}--}}} & \multicolumn{2}{c||}{\textbf{\analyze{}--}} & \multicolumn{3}{c}{\textbf{{\evaluate{}--}}}\\ 
        {} & \multicolumn{2}{c||}{\textbf{{\analyze{}}}} & \multicolumn{2}{c||}{\textbf{\evaluate{}}} & \multicolumn{3}{c}{\textbf{{\create{}}}}\\
        & {Solution} & {Solution}  & {Code} & {Code} & {\textsc{Avatar}} & {\textsc{Goal}} & {\textsc{Wall}} \\
        & {checking} & {tracing}  & {debugging} & {equivalence} & {design} & {design} & {design} \\
        \midrule
        {Basic moves and turns} & {Q01} & {} &{Q08} & {} & {Q15} &  {} & {} \\[0.28cm]
        \hline
        \DSLRepeat\textcode{\{\}} & {Q02} & {Q05} & {} & {Q12} & {Q16} & {} & {} \\[0.18cm]
        \DSLRepeatUntil\textcode{\{\}} & {} & {Q06} & {Q09} & {} & {Q17} & {Q18} & {} \\[0.18cm]
        \hline
        \DSLRepeatUntil\textcode{\{}\DSLIf\textcode{\}} & {Q07} & {} & {Q10} & {} & {} & {Q19} & {} \\[0.18cm]
        \DSLRepeatUntil\textcode{\{}\DSLIfElse\textcode{\}} & {Q04} & {} & {Q11} & {Q14} & {} & {} & {Q20, Q21} \\[0.18cm]
        \hline
        \DSLRepeat\textcode{\{}\DSLRepeat\textcode{\}} & {} & {} & {} & {Q13} & {} & {} & {} \\
        \DSLRepeat\textcode{\{}\DSLIf\textcode{\}} & {Q03} & {} & {} & {} & {} & {} & {}\\[0.18cm]
        \bottomrule
        \end{tabular}
    }
    \subcaption{Our Test: Distribution of Items}
    \label{fig1:test_questions.dist}
    \end{subfigure}
    \begin{subfigure}{0.49\textwidth}
        \centering
        \setlength{\fboxsep}{0.05pt}\fbox{\includegraphics[width=0.92\linewidth]{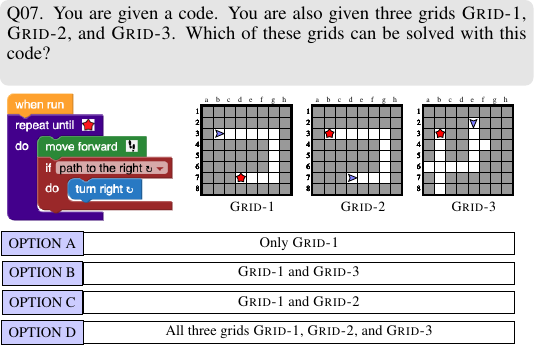}}
        \subcaption{Q07. Solution checking}
        \label{fig1:test_questions.s}
    \end{subfigure}
    \\
    \begin{subfigure}{0.49\textwidth}
        \centering
        \setlength{\fboxsep}{0.05pt}\fbox{\includegraphics[width=0.92\textwidth]{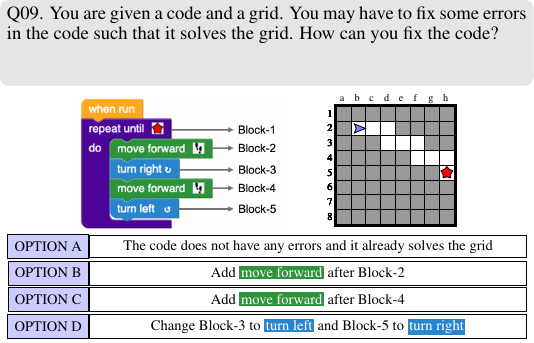}}
        \subcaption{Q09. Code debugging}
        \label{fig1:test_questions.db1}
        \end{subfigure}
    %
    \begin{subfigure}{0.49\textwidth}
        \centering
        \setlength{\fboxsep}{0.05pt}\fbox{\includegraphics[width=0.92\textwidth]{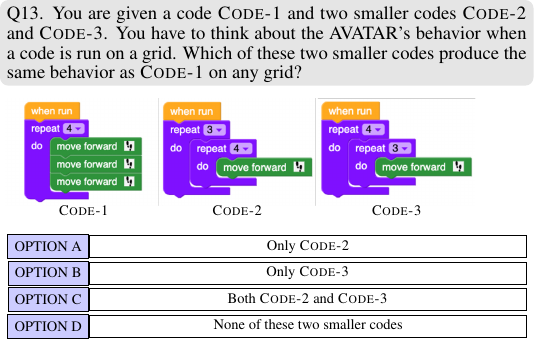}}
        \subcaption{Q13. Code equivalence}
        \label{fig1:test_questions.db2}
    \end{subfigure}
    \\
    \begin{subfigure}{0.49\textwidth}
        \centering
        \setlength{\fboxsep}{0.05pt}\fbox{\includegraphics[width=0.92\textwidth]{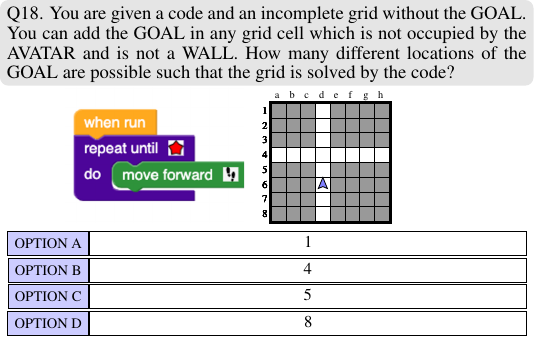}}
        \subcaption{Q18. \textsc{Goal} design}
        \label{fig1:test_questions.ds1}
    \end{subfigure}
    %
    \begin{subfigure}{0.49\textwidth}
        \centering
        \setlength{\fboxsep}{0.05pt}\fbox{\includegraphics[width=0.92\textwidth]{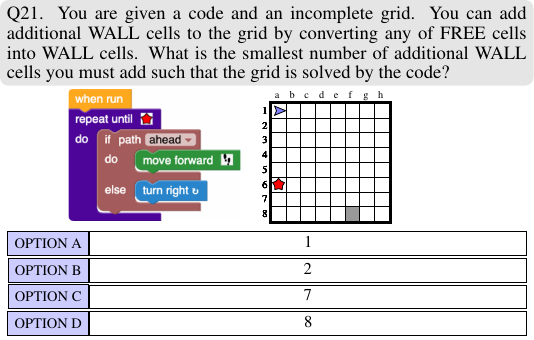}}
        \subcaption{Q21. \textsc{Wall} design}
        \label{fig1:test_questions.ds2}
    \end{subfigure}
    %
    \caption{\looseness-1(a) shows the distribution of test items w.r.t to CT and problem-solving concepts and Bloom's cognitive levels. (b)--(f) are examples of five items from \ourtest{}. 
    These items are grounded in the domain of \emph{Hour of Code: Maze Challenge} (\hocmaze{})~\citep{hourofcode_maze}, which can be found at {\textcolor{urlcolor}{studio.code.org/s/hourofcode}}. \hocmaze{} domain comprises elementary block-based visual programming tasks where one has to write a solution code that would navigate the \textsc{Avatar} (blue dart) to the \textsc{Goal} (red star) without crashing into \textsc{Wall}s (gray grid cells). We encourage the reader to attempt these items; all $21$ test items from \ourtest{} along with their answers are provided in the appendix.}
    \label{fig1:test_questions}
\end{figure*}

To this end, we have developed a novel test for grades $3$-$7$, \ourtest{}, that focuses on the three higher cognitive levels of Bloom's Taxonomy, i.e., \analyze{}, \evaluate{}, and \create{}~\citep{bloom2020taxonomy}. It comprises a diverse set of multiple-choice items spanning all three higher cognitive levels, including the highest level of \create{}. Figure~\ref{fig1:test_questions} illustrates the diversity of items covered by \ourtest{}. Further details of the development of \ourtest{} are presented in Section~\ref{sec.ourtest}. In this paper, our objective is to validate \ourtest{} with students from grades $3$--$7$, and report on its psychometric properties. Specifically, we center the analysis around the following research questions: (1) \textbf{RQ1}: How is the internal structure of \ourtest{} organized w.r.t. item categories pertaining to Bloom's higher cognitive levels? (2) \textbf{RQ2}: What is the reliability of \ourtest{} w.r.t. consistency of its items? (3) \textbf{RQ3}: How does performance on \ourtest{} correlate with performance on real-world programming platforms and students' prior programming experience?


\begin{table*}
    \caption{Categorization of different CT assessments proposed in recent works. The first column shows the specific CT Assessment. The next three columns, \apply{}-\analyze{}, \analyze{}-\evaluate{}, and \evaluate{}-\create{}, classify the assessment based on these different cognitive levels of Bloom's Taxonomy where ``\yestick{}'' implies presence of the levels and ``\notick{}'' implies absence of the levels. The ``Grade'' column refers to the intended grades (age group) for the test. The ``Validity'' column refers to three dimensions across which the test was validated, including (i) ``Student'': test items validated with students; (ii) ``Expert'': test items validated with experts; (iii) ``Convergent'': test validated w.r.t. performance on another test/course. Finally, the ``Domain'' column shows the domain on which the items in the test were designed. Further details are presented in Section~\ref{sec.relatedwork}.}
    \label{fig2:relatedwork}
    \vspace{-3mm} 
    \centering{
    \scalebox{0.81}{
    \setlength\tabcolsep{4.8pt}
    \renewcommand{\arraystretch}{1.22}  
    \begin{tabular}{l | c | c | c | c | ccc | c }
    \toprule 
    \vspace{-1.50mm}
    {\textbf{CT Assessment and Tests}} & \textbf{\apply{}--} & \textbf{\analyze{}--} & \textbf{\evaluate{}--} & \textbf{Grades} & \multicolumn{3}{c|}{\textbf{Validity:}} & \textbf{Domain} \\
    {} & \textbf{\analyze{}} & \textbf{\evaluate{}} & \textbf{\create{}} & {} & {Student} & {Expert} & {Convergent} & {} \\
    \midrule
    \ {\cellcolor{green!15}}{\textbf{\ourtest{}} {(this paper)}} & {\cellcolor{green!15}}\yestick{} & {\cellcolor{green!15}}\yestick{} & {\cellcolor{green!15}}\yestick{} & {\cellcolor{green!15}}$3$-$7$ & {\cellcolor{green!15}}\yestick{} & {\cellcolor{green!15}}\yestick{} & {\cellcolor{green!15}}\yestick{} &  {\cellcolor{green!15}}block-based visual programming \\
    \hline
    \ cCTt~\citep{DBLP:journals/corr/abs-2203-05980} & \yestick{} & \yestick{} & \notick{} & $3$-$4$ & \yestick{} & \yestick{} & \notick{} & {block-based visual programming} \\
    {\ TechCheck}~\citep{relkin2020techcheck} & \yestick{} & \yestick{} & \notick{} & K-$4$ & \yestick{} & \yestick{} & \yestick{} & {everyday scenarios} \\
    {\ CT-Test}~\citep{gonzalez2015computational, DBLP:journals/chb/Roman-GonzalezP17, roman2019combining} & \yestick{} & \yestick{} & \notick{} & $6$-$8$ &  \yestick{} & \yestick{} & \yestick{} & {block-based visual programming}  \\
    \vspace{-1.9mm}    
    {Gane et al. 2021}~\citep{DBLP:journals/csedu/GaneIEYLP21} & \yestick{} & \yestick{} & \notick{} & $3$-$4$ & \yestick{} & \yestick{} & \notick{} & {block-based visual programming} \\
     {} & {} & {} & {} & {} & {} & {} & {} & and everyday scenarios \\
    \vspace{-1.9mm}
    {Chen et al. 2017}~\citep{DBLP:journals/ce/ChenSBJHE17} & \yestick{} & \yestick{} & \notick{} & $5$ &  \yestick{} & \yestick{} & \notick{} & {robotics programming} \\
    {} & {} & {} & {} & {} & {} & & {} & {} and everyday scenarios \\
    \vspace{-1.9mm}
    {ACES}~\citep{DBLP:conf/sigcse/ParkerKSFKRW21}  & \yestick{} & \yestick{} & \notick{} & $3$-$5$ & \yestick{} & \notick{} & \notick{} & block-based visual programming \\
    {}  & {} & {} & {} & {} & {} & {} & {} & {} and everyday scenarios  \\
    \hline
    \vspace{-1.9mm}    
    {CTC}~\citep{DBLP:journals/jeric/Lai22} & \yestick{} & \yestick{} & \notick{} & $8$-$12$ & \yestick{} & \yestick{} & \yestick{} & block-based visual programming  \\
    {} & {} & {} & {} & {} & {} & {} & {} & and real-world problem-solving \\
    \vspace{-1.9mm}
    {Commutative Assessment}~\citep{DBLP:conf/icer/WeintropW15} & \yestick{} & \yestick{} & \notick{} & $8$-$12$ & \yestick{} & \yestick{} & \notick{} & {block-based visual programming} \\
    {} & {} & {} & {} & {} & {} & {} & {} & and text-based visual programming \\
    {\ Mühling et al. 2015}~\citep{DBLP:conf/wipsce/MuhlingRH15}  & \yestick{} & \yestick{} & \notick{} & $8$-$10$ & \yestick{} & \notick{} & \notick{} & {similar to Karel programming \cite{pattis1981karel}} \\
    \hline
    \vspace{-1.9mm}    
    {PSIv1}~\citep{DBLP:conf/sigcse/BockmonB23}& \yestick{} & \notick{} & \notick{} & $13$-$14$ & \yestick{} & \yestick{} & \yestick{} & {text-based programming} \\
    {} & {} & {} & {} & {(college)} & {} & {} & {} & {} \\   
    \bottomrule
    \end{tabular}
    } 
    } 
\end{table*}   



\section{Related Work}\label{sec.relatedwork}
Prior work has proposed several CT assessments and their categorizations based on their format including the following~\citep{roman2019combining, DBLP:journals/corr/abs-2203-05980}: (a) \emph{portfolios}, which are project-based programming assessments; (b) \emph{interviews}, which are used in conjunction with portfolios to gain insights into students' thinking process; (c) \emph{summative assessments}, which are long-format answer type questions to measure CT specifically in the context of a particular domain; (d) \emph{multi-choice diagnostic tests}, which measure CT aptitude and may be administered in both pretest and posttest conditions. As mentioned in Section~\ref{sec.intro}, we focus on multi-choice CT tests due to their practicality and scalability. Table~\ref{fig2:relatedwork} presents several different multi-choice diagnostic tests proposed in the literature, viewed through the lens of Bloom's taxonomy~\citep{roman2019combining}. Specifically, we classify them based on their coverage of the higher cognitive levels of the taxonomy (\apply{}, \analyze{}, \evaluate{}, and \create{}). 

These tests cater to students from different school years, starting from kindergarten (K) through the early years of college. Next, we describe three representative assessments in different years. The \emph{competent Computational Thinking test} (cCTt)~\citep{DBLP:journals/corr/abs-2203-05980} was proposed recently in 2022 for students in grades $3$--$4$. The test comprised items that only required finding solution codes or completing a given solution code. These types of items invoke students' \apply{}, \analyze{}, and \evaluate{} cognitive levels. The \emph{Computational Thinking Challenge} (CTC)~\citep{DBLP:journals/jeric/Lai22} was proposed in 2021 for students in grades $9$--$12$. The test contains programming items in the form of Parsons problems~\citep{DBLP:conf/icer/ZhiCBP19}, solution-finding multi-choice items, and general items on real-world problem-solving. The items in CTC also cover all cognitive levels except the \create{} level. Finally, \emph{Placement Skill Inventory v1} (PSIv1)~\citep{DBLP:conf/sigcse/BockmonB23}
was also proposed recently in 2022 for college students as a placement test. The test contains multi-choice theoretical items on programming and covers only the \apply{} and \analyze{} levels of Bloom's taxonomy. Contrary to these tests, \ourtest{} contains items that require synthesizing new problem instances to verify the correctness of a proposed solution. These items in \ourtest{} are intended to cover the Bloom's \create{} cognitive level. \ourtest{} is developed for students in grades $3$--$7$.
%

Table~\ref{fig2:relatedwork} also shows different domains in which CT is measured in these tests. For grades K--$8$, the most popular setting is block-based visual programming, likely because of the low syntax overhead of the domains and the ease of measuring CT concepts such as conditionals, loops, and sequences~\citep{ DBLP:journals/chb/Roman-GonzalezP17, DBLP:conf/icer/WeintropW15, DBLP:journals/corr/abs-2203-05980,gonzalez2015computational}. Beyond block-based programming domains, several CT tests also utilize real-world settings, including everyday-scenarios (e.g., a scenario related to seating arrangements in a gathering)~\citep{DBLP:conf/sigcse/ParkerKSFKRW21}, robotics~\citep{DBLP:journals/ce/ChenSBJHE17}, and real-world problem-solving (e.g., a problem related to route planning in a city)~\citep{DBLP:journals/jeric/Lai22}. The advantage of these real-world settings and domains is their administration with minimal domain knowledge, thus making them suitable pretest and posttest candidates. \ourtest{} is based on the block-based visual programming domain.


\begin{figure*}
\centering{
    \begin{subfigure}{0.3\textwidth}
    \centering
    \includegraphics[width=0.90\textwidth]{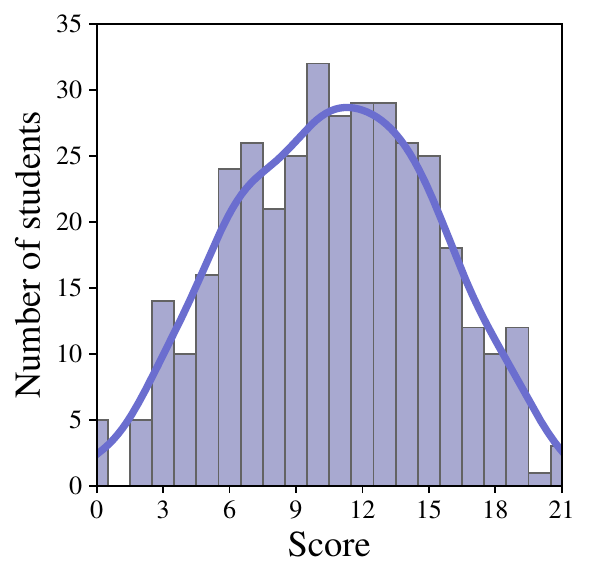}
    \subcaption{Overall Distribution of Test Scores}
    \label{fig2:overview_results.cumulativedist}
    \end{subfigure}
    %
    \begin{subfigure}{0.305\textwidth}
    \centering
    \includegraphics[width=0.90\textwidth]{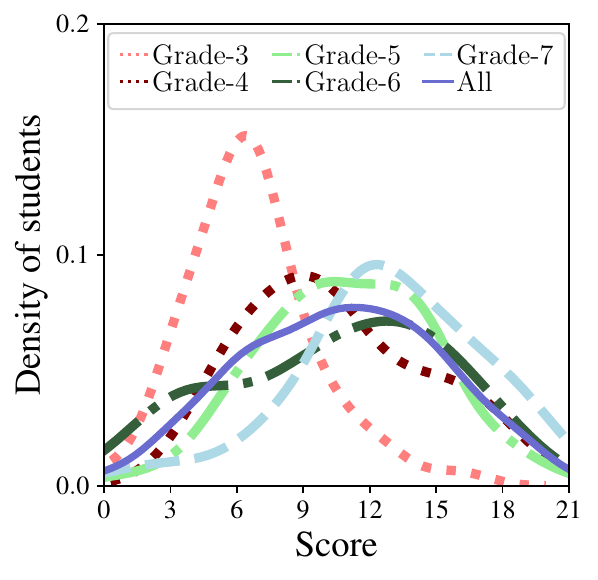}
    \subcaption{Grade-wise Distribution of Test Scores}
    \label{fig2:overview_results.gradedist}
    \end{subfigure}
    %
    \begin{subfigure}{0.38\textwidth}
    \centering
    \includegraphics[width=0.90\textwidth]{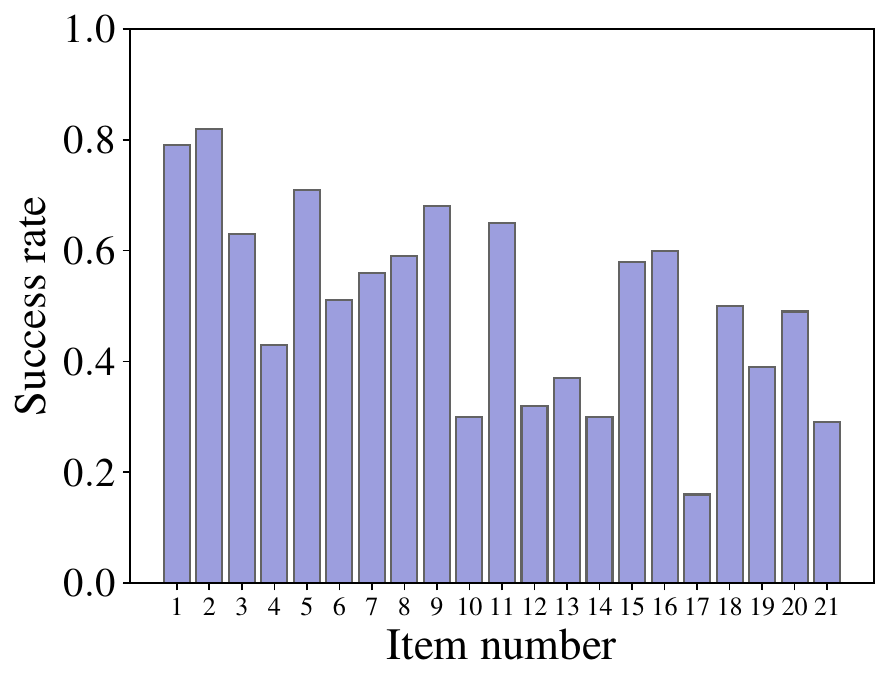}
    \subcaption{Success Rate per Item}
    \label{fig2:overview_results.quesdist}
    \end{subfigure}
}
\caption{An overview of the performance of students on \ourtest{}. (a) overall distribution of \ourtest{} scores across all $371$ students; (b) distribution of \ourtest{} score per grade; (c) success rate of students for each item in \ourtest{}. Details are in Section~\ref{sec.setup}.}
\label{fig2:overview_results}
\end{figure*}

Finally, an important aspect of developing such CT assessments is their validation and reliability~\citep{DBLP:journals/corr/abs-2203-05980}. Generally, CT assessments are validated using three methods: (a) with students in specific grades for which the assessment was designed; (b) with expert feedback; (c) w.r.t. another test or performance in a course (i.e., convergent validity). For a well-rounded evaluation, it is advisable to explore all three validation methods~\citep{roman2019combining, DBLP:journals/corr/abs-2203-05980}. As shown in Table~\ref{fig2:relatedwork}, most tests are validated with students, while some are refined by experts. However, the incorporation of convergent validity is less common. \ourtest{} is validated using all three methods.


\section{Our Test: \ourtestheader{}}\label{sec.ourtest}
The development of \ourtest{} is centered around the higher cognitive levels of Bloom's taxonomy: \analyze{}, \evaluate{}, and \create{}. The test contains items grounded in the domain of block-based visual programming. Specifically, we consider the popular block-based visual programming domain of \emph{Hour of Code: Maze Challenge}~\citep{hourofcode_maze} by code.org~\citep{codeorg}. We picked this domain as it encapsulates important CT and problem-solving concepts of conditionals, loops, and sequences, within the simplicity of the block-based structure.
Students can attempt tasks in this domain with a simple description of the constructs and task, as discussed in the caption of Figure~\ref{fig1:test_questions}. Next, we describe the 
items in \ourtest{} which are divided into the following three categories based on Bloom's higher cognitive levels:
\begin{itemize}[leftmargin=1em]
    \item \apply{}--\analyze{}: This category comprises items either on finding a solution code of a given task or reasoning about the trace of a given solution code on one or more visual grids. They are based on the \apply{} and \analyze{} levels of Bloom's taxonomy, as they require applying CT concepts and analyzing code traces. These items are typically the most common type of items included in several CT tests~\citep{DBLP:journals/corr/abs-2203-05980, DBLP:conf/sigcse/ParkerKSFKRW21}.
    \item \analyze{}--\evaluate{}: This category comprises items that require reasoning about errors in candidate solution codes of a task and evaluating the equivalence of different codes for a given task. They are based on the \analyze{} and \evaluate{} levels of Bloom's taxonomy. Several CT assessments also include these types of debugging items~\citep{DBLP:journals/jeric/Lai22, gonzalez2015computational}.
    \item \evaluate{}--\create{}: This category comprises items that require reasoning about the design of task grids for given solution codes. They are based on the \evaluate{} and \create{} levels of Bloom's taxonomy, as they involve synthesizing components of visual grids such as \textsc{Avatar}, \textsc{Goal}, and \textsc{Wall}. These items are unique to \ourtest{} and capture the open-ended nature of task design, such as counting several possible task configurations to satisfy a given solution code (see items Q18 and Q21 in Figure~\ref{fig1:test_questions}).
\end{itemize}

\looseness-1In the process of developing \ourtest{}, we consulted with  CS educators and researchers with expertise in using CT tools for K-12 education. Five experts provided feedback on an initial version of the test in terms of items' suitability and difficulty. Furthermore, we asked six students (not part of the study) from grades $3$ to $6$ to attempt a version of the test while recording their thought processes via think-aloud methods. The responses of these students allowed us to further refine the phrasing and structure of the items. Ultimately, the final version of \ourtest{} contained $21$ single-correct multiple-choice items to be completed in one $45$-minute lesson. The items were divided equally across three categories based on Bloom's cognitive levels, i.e., each category comprised $7$ items. The three categories are henceforth referred to as \ourtestone{}, \ourtesttwo{}, and \ourtestthree{}, respectively. Figure~\ref{fig1:test_questions} presents the breakdown of all our $21$ test items and also illustrates $5$ items from \ourtest{}. All $21$ test items with their answers are presented in Appendices~\ref{sec.appendix.ace} and~\ref{sec.appendix.ace.solution}.

%



\begin{figure*}
\centering{
    \begin{subfigure}{0.35\textwidth}
    \centering
    \includegraphics[width=0.90\textwidth]{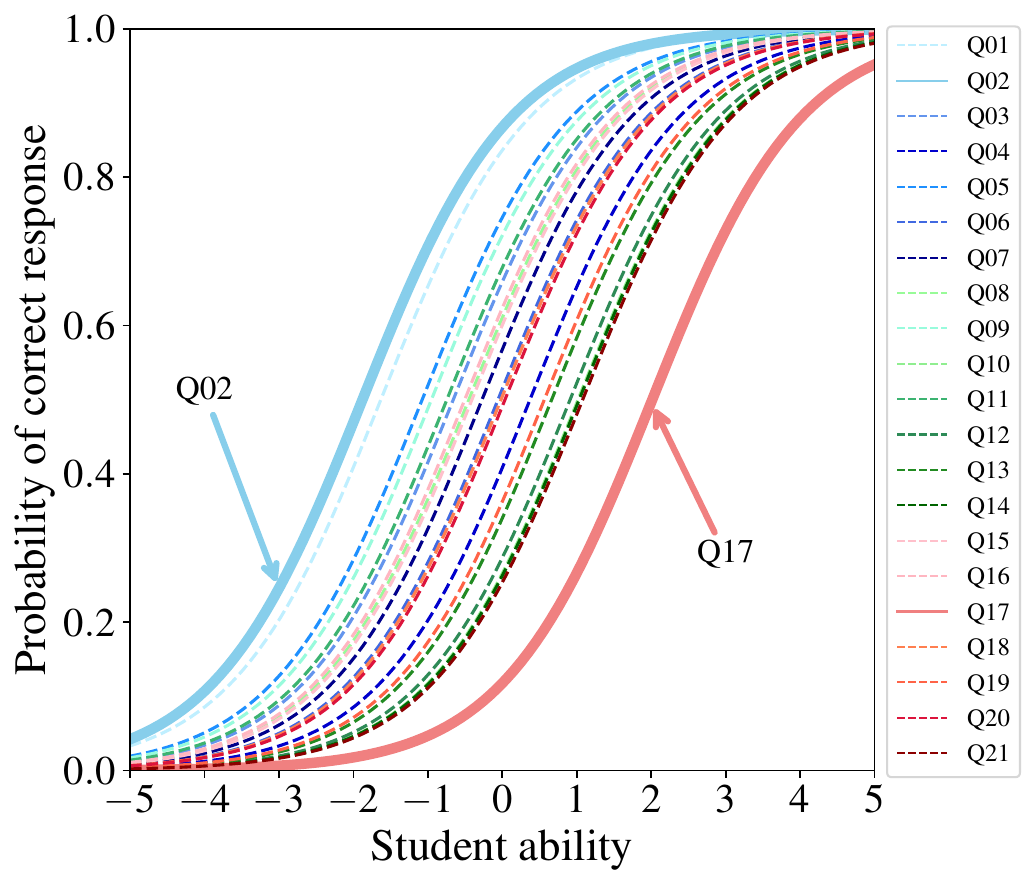}
    \subcaption{Item Characteristic Curves}
    \label{fig3:irl_corr_results_icc}
    \end{subfigure}
    %
    \begin{subfigure}{0.58\textwidth}
    \centering
    \includegraphics[width=0.90\textwidth]{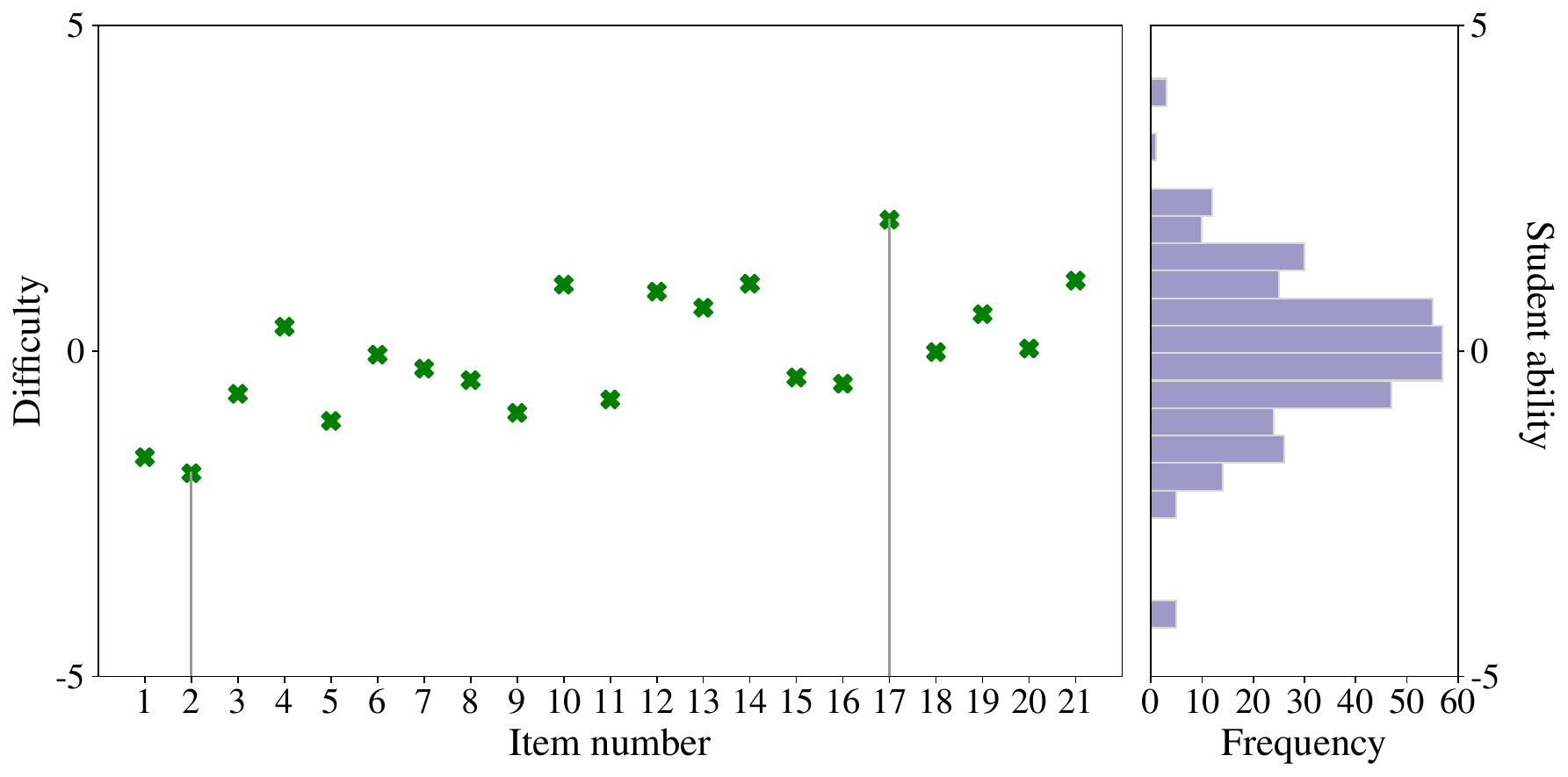}
    \subcaption{Wright Map: Item and Student Distribution}
    \label{fig3:irl_corr_results_wrightmap}
    \end{subfigure}
}
\caption{Results from a 1-parameter Rasch model~\citep{rasch1993probabilistic} on the \ourtest{} items and student scores. (a) Item characteristic curve for each item in \ourtest{} and (b) Wright map corresponding to our student population.}
\label{fig3:irl_corr_results}
\end{figure*}

\section{Study and Descriptive Statistics}\label{sec.setup}
In this section, we provide details of the data collection process for \ourtest{}'s psychometric evaluation.

\subsection{Two-Phase Data Collection Process}\label{sec.setup.survey}
The study to evaluate the psychometric properties of \ourtest{} was planned in two phases, spread across two weeks. The first phase was intended to familiarize students with the block-based visual programming domain of \emph{Hour of Code: Maze Challenge} (\hocmaze{})~\citep{hourofcode_maze} by code.org~\citep{codeorg}, and introduce them to basic programming concepts. Additionally, it would serve as a baseline to correlate students' performance w.r.t. \ourtest{}, and measure the convergent validity of \ourtest{}. In the second phase of the study, students would take the \ourtest{} test. This two-phase study design ensured that students would have enough focus on each study component as well as a time gap between domain familiarity and the actual test.

We obtained an Ethical Review Board approval from the Ethics Committee of Tallinn University before conducting the study. The study was conducted in Estonia, where a random selection of $10$ schools was pooled from $11$ out of $15$ counties. Participation in the study was voluntary for both teachers and students. Next, we outline the details of each phase of the study.

\looseness-1The data collection process was conducted in May 2023. During both phases, students received usernames to ensure anonymity throughout the study. The first phase of data collection included one $45$-minute lesson during which the students filled in a short background questionnaire in Google Forms (about $5$ minutes) and then solved $20$ tasks from \hocmaze{} (about $40$ minutes). We hosted these $20$ tasks on a separate platform created for the study to enable the collection of students' performance data on these tasks. Students were allowed multiple attempts to solve each task and could score a maximum of $20$ points, i.e., $1$ point per task. Henceforth, we refer to students' performance in this phase as their \hocmaze{} score. The second phase took place one week later and involved a $45$-minute lesson during which students took the \ourtest{} test. The test was administered through a Qualtrics survey. Students could score a maximum of $21$ points, i.e., $1$ point per item. Henceforth, we refer to students' performance in this phase as their \ourtest{} score.

\subsection{Participants and Descriptive Statistics}\label{sec.setup.descriptive_stats}
A total of $371$ students participated in the study, distributed across five different grades as follows: $n=51$ from grade $3$, $n=34$ from grade $4$, $n=114$ from grade $5$, $n=81$ from grade $6$, and $n=91$ from grade $7$. The participants' age varied from $9$ to $15$ years. The distribution of students w.r.t. gender was as follows: $48\%$ girls and $52\%$ boys. Regarding the students' prior programming experience, their reported duration of programming experience was as follows: $40\%$ reported no programming experience, $22\%$ reported at least $1$ year, $15\%$ reported at least $2$ years, $13\%$ reported at least $3$ years, and $10\%$ reported between $4$ to $6$ years. Among the $60\%$ of participants ($n=223$) having programming experience, they reported the following sources of gaining programming experience (multiple sources could be selected): $77\%$ studied programming in school lessons, $31\%$ studied programming in after-school lessons, and $14\%$ studied programming independently at home. 
Figure~\ref{fig2:overview_results} summarizes the students' performance on \ourtest{}. The average score of all $371$ students is $10.69$. The distribution of the scores across all participants is shown in Figure~\ref{fig2:overview_results.cumulativedist}. Grade $7$ averaged the highest score of $12.84$, while grade $3$ averaged the lowest score of $6.67$. Grades $4$, $5$, and $6$ had similar average scores of around $10.5$. Figure~\ref{fig2:overview_results.gradedist} shows the distribution of scores per grade, and Figure~\ref{fig2:overview_results.quesdist} shows the success rate per item. Overall, the success rate of \ourtestone{} was higher than those of \ourtesttwo{} and \ourtestthree{}. More concretely, the average score of all $371$ students when measured per category (i.e., points on $7$ items in each category) was as follows: $4.45$ on \ourtestone{}, $3.22$ on \ourtesttwo{}, and $3.02$ on \ourtestthree{}.

%



\section{Results and Discussion}\label{sec.results}
In this section, we discuss the results of the study centered around the research questions (RQs) introduced in Section~\ref{sec.intro}.

\subsection{RQ1: Internal Structure of \ourtestheader{}}\label{sec.results.rq1}
We assess the internal structure of \ourtest{} w.r.t. its three item categories (\ourtestone{}, \ourtesttwo{}, \ourtestthree{}) as its underlying factors using Confirmatory Factor Analysis (CFA), a standard method in quantitative test analysis~\citep{distefano2005using, DBLP:journals/jeric/Lai22}. CFA determines whether the structure of \ourtest{} scores aligns with the three item categories as three factors. Specifically, CFA provides the Root Mean Square Error of Approximation (RMSEA) as the goodness of fit for statistical models. RMSEA values between $0$ and $0.01$ indicate excellent fit, and values up to $0.05$ indicate good fit. Additionally, CFA provides the Comparative Fit Index (CFI) and Tucker-Lewis Index (TLI). CFI measures how well our three-factor model fits the observed data on test scores compared to an independent-item model with $21$ items as factors. TLI also measures model fit by accounting for model simplicity. CFI and TLI values greater than $0.90$ indicate a good fit. 

Our test presented a significant model with good fit statistics ($p<0.01$; RMSEA=$0.0275$; CFI=$0.945$; TLI=$0.938$). This analysis also highlighted a potential issue with item Q17, as the Standardized Estimate for \emph{CFA's factor loading} of this item seemed problematic with a value of $-0.111$ ($p = 0.059$). Nevertheless, the model fit did not improve significantly after removing item Q17 from the data; new statistics without Q17 ($p<0.01$; RMSEA=$0.0278$; CFI=$0.949$; TLI=$0.942$) are similar to statistics reported above without any significant difference. We discuss this item further as part of RQ2.


\subsection{RQ2: Reliability of \ourtestheader{}}\label{sec.results.rq2}
Next, we determine the reliability of \ourtest{}, i.e., a measure of its ability to produce consistent and stable results over repeated administrations (a higher value being better). One  standard way to measure this is through the Cronbach alpha value~\citep{tavakol2011making, DBLP:journals/jeric/Lai22} that reflects the average inter-item correlations in a test. Another method is the reliability of student ability estimates obtained from Item Response Theory (IRT). In our study, we apply IRT analysis on students' responses to \ourtest{} and fit a 1-parameter logic Rasch model (1-PL IRT)~\citep{rasch1993probabilistic}. The model estimates the per-item difficulties and students' abilities, and provides the reliability of these estimates.

The overall reliability for our test was \emph{good} with a Cronbach alpha value of $0.813$. Among the three item categories, Cronbach alpha was $0.622$ for \ourtestone{}, $0.562$ for \ourtesttwo{}, and $0.625$ for \ourtestthree{}. Figure~\ref{fig3:irl_corr_results_icc} shows the 1-PL IRT item characteristic curves for all items; we find that Q02 is the easiest and Q17 is the hardest \ourtest{} item. Figure~\ref{fig3:irl_corr_results_wrightmap} illustrates the difficulty of items as well as the estimated ability of students' in our population. The 1-PL IRT Person reliability value for all $21$ items is $0.790$ (with $p < 0.01$). 

\begin{figure}[t!]
\setlength{\fboxsep}{0.05pt}\fbox{\includegraphics[width=0.42\textwidth]{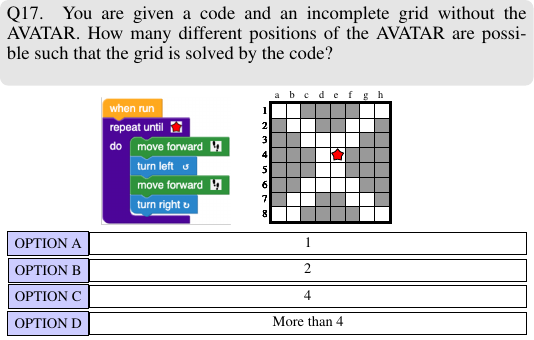}}
\caption{Q17. \textsc{Avatar} design}
\label{fig4:test_questions.ds03}
\end{figure}

Next, we discuss the potentially problematic item Q17 shown in Figure~\ref{fig4:test_questions.ds03}. We find that its exclusion from the model doesn't significantly improve the IRT Person reliability. One possible reason Q17 prompted incorrect responses is that it was the first item in \ourtest{} requiring enumeration of all possible \textsc{Avatar} locations. However, students adapted to similar formats in subsequent items (e.g., Q18 and Q21 in Figure~\ref{fig1:test_questions}). Prior work confirms that varying response formats can cause deviations~\cite {DBLP:journals/tkl/GuggemosSR23}. A possible revision of item Q17 could be simplifying the visual grid to reduce its complexity.

\subsection{RQ3: Correlating \ourtestheader{} scores}\label{sec.results.rq3}
\looseness-1We measure the convergent validity of \ourtest{} w.r.t. \hocmaze{} scores. Additionally, we also measure the correlation of the three \ourtest{} categories with both \hocmaze{} scores as well as overall \ourtest{} scores. Finally, we measure the influence of extrinsic factors such as prior programming experience on \ourtest{} scores. To measure all these correlations, we perform standard Pearson's correlation analysis between each of these features on data from our entire student population~\citep{sedgwick2012pearson, DBLP:journals/jeric/Lai22}. High positive values of Pearson's correlation coefficient, $r$, indicate a strong positive correlation. 

\begin{figure}[t!]
    \scalebox{0.83}{
    \includegraphics[width=0.45\textwidth]{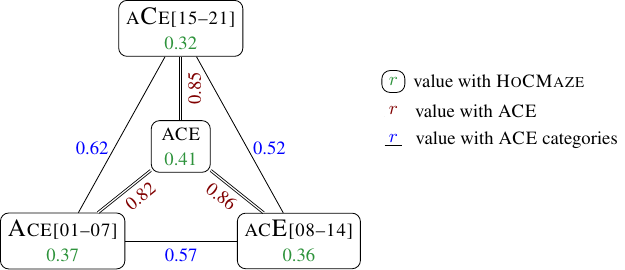}}
    \caption{Pearson's correlation coefficient, $r$, between \ourtest{} and \hocmaze{}, between \ourtest{} and its categories, and between each category. All values are significant with $p<0.001$.}
    \label{fig6:pearsons_corr}
\end{figure}

\looseness-1The correlation results for \ourtest{} with \hocmaze{} and item categories of \ourtest{} are shown in Figure~\ref{fig6:pearsons_corr}. For instance, the Pearson correlation between \ourtest{} scores and \hocmaze{} scores w.r.t. all students is $r=0.41, p < 0.001$, confirming the convergent validity of \ourtest{}. All item categories were (significantly) positively correlated with \ourtest{} and also have high positive inter-category correlations.

In terms of the effect of prior programming experience on \ourtest{}, we observed a significant positive correlation with both the student's year of study ($r=0.358$, $p<0.01$) and age ($r=0.359$, $p<0.01$).
Our result aligns with prior work~\citep{DBLP:conf/icer/SwidanHS18} indicating that participants' developmental factors (e.g., reading skills, abstract thinking) can impact test performance. In our student population, varying programming exposure due to elective programming courses influenced prior programming skills. Analyzing this further, we discovered that students who took after-school programming classes outperformed those who did not on \ourtest{} ($p < 0.05$, w.r.t. $t$-test~\citep{welch1947generalization}).
%

\subsection{Limitations}\label{sec.results.limitations}
Next, we discuss a few limitations of our current study. Firstly, in this study, we evaluated the convergent validity of \ourtest{} w.r.t. \hocmaze{} scores. However, it would be more informative to evaluate \ourtest{} w.r.t. other types of assessments, such as portfolios, which specifically consider \create{} cognitive level. Moreover, it would be interesting to evaluate the convergent validity of \ourtest{} w.r.t. students' performance in other subjects involving CT. Secondly, grade $3$ did not present a significant correlation between \ourtest{} and \hocmaze{} scores (Pearson's $r=0.068; p=0.633$), possibly because of difficulties with text comprehension of the item descriptions. Hence, refining the presentation of items could be beneficial for this age group. Finally, we presented the test items in a fixed order, which might have affected students' performance on specific items such as Q17. Implementing a randomized order of the test items within each category could be a way to address this limitation.



\section{Conclusion and Future Work}
\label{sec.conclusion}
We developed a new test, \ourtest{}, to assess CT and problem-solving skills, focusing on higher levels of Bloom's taxonomy, including \create{}. We capture this level through a novel category of items that go beyond solution finding or debugging and consider task design. In this paper, we studied the psychometric properties of \ourtest{}, and our results confirm \ourtest{}'s reliability and validity. There are several exciting directions for future work. Firstly, we can extend the framework of items to develop tests with more advanced programming constructs, such as variables/functions suitable for higher grades. Secondly, while we studied the utility of items in \ourtest{} for CT assessments, these items could also be incorporated as part of the curriculum to teach students richer CT and problem-solving skills such as problem design and test-case creation.



\section{Acknowledgements}
We would like to thank the teachers and students in Estonia for their participation in the study. We thank Danial Hoosyar and the Ethics Committee of Tallinn University, Estonia, for reviewing and providing the Ethical Review Board (ERB) approval for the study. Ahana Ghosh acknowledges support from Microsoft Research through its PhD Scholarship Programme. Funded/Cofunded by the European Union (ERC, TOPS, 101039090). Views and opinions expressed are however those of the author(s) only and do not necessarily reflect those of the European Union or the European Research Council. Neither the European Union nor the granting authority can be held responsible for them.


\balance
\bibliographystyle{unsrt}
\bibliography{main}

\begin{thebibliography}{10}

\bibitem{wing2006computational}
Jeannette~M Wing.
\newblock {C}omputational {T}hinking.
\newblock {\em {C}ommunications of the {ACM}}, 2006.

\bibitem{DBLP:journals/jeric/GroverBBEDS17}
Shuchi Grover, Satabdi Basu, Marie~A. Bienkowski, Michael Eagle, Nicholas
  Diana, and John~C. Stamper.
\newblock {A} {F}ramework for {U}sing {H}ypothesis-{D}riven {A}pproaches to
  {S}upport {D}ata-{D}riven {L}earning {A}nalytics in {M}easuring
  {C}omputational {T}hinking in {B}lock-{B}ased {P}rogramming {E}nvironments.
\newblock {\em Transactions on Computing Education}, 2017.

\bibitem{chevalier2020fostering}
Morgane Chevalier, Christian Giang, Alberto Piatti, and Francesco Mondada.
\newblock {F}ostering {C}omputational {T}hinking {T}hrough {E}ducational
  {R}obotics: {A} {M}odel for {C}reative {C}omputational {P}roblem {S}olving.
\newblock {\em International Journal of STEM Education}, 2020.

\bibitem{grover2013computational}
Shuchi Grover and Roy Pea.
\newblock {C}omputational {T}hinking in {K}--12: {A} {R}eview of the {S}tate of
  the {F}ield.
\newblock {\em Educational Researcher}, 42(1):38--43, 2013.

\bibitem{DBLP:conf/chi/LoksaKJOMB16}
Dastyni Loksa, Amy~J. Ko, Will Jernigan, Alannah Oleson, Christopher~J. Mendez,
  and Margaret~M. Burnett.
\newblock {P}rogramming, {P}roblem {S}olving, and {S}elf-{A}wareness: {E}ffects
  of {E}xplicit {G}uidance.
\newblock In {\em {CHI}}, 2016.

\bibitem{DBLP:journals/ijcses/LockwoodM18}
James Lockwood and Aidan Mooney.
\newblock {C}omputational {T}hinking in {S}econdary {E}ducation: {W}here {D}oes
  it {F}it? {A} {S}ystematic {L}iterary {R}eview.
\newblock {\em International Journal of Computer Science Education in Schools},
  2018.

\bibitem{DBLP:journals/corr/abs-2203-05980}
Laila~El Hamamsy, Mar{\'{\i}}a Zapata{-}C{\'{a}}ceres, Estefan{\'{\i}}a
  Mart{\'{\i}}n{-}Barroso, Francesco Mondada, Jessica Dehler{-}Zufferey, and
  Barbara Bruno.
\newblock {T}he competent {C}omputational {T}hinking test (c{CT}t):
  {D}evelopment and {V}alidation of an {U}nplugged {C}omputational {T}hinking
  {T}est for {U}pper {P}rimary {S}chool.
\newblock {\em CoRR}, abs/2203.05980, 2022.

\bibitem{DBLP:conf/educon/Zapata-CaceresM20}
Mar{\'{\i}}a Zapata{-}C{\'{a}}ceres, Estefan{\'{\i}}a Mart{\'{\i}}n{-}Barroso,
  and Marcos Rom{\'{a}}n{-}Gonz{\'{a}}lez.
\newblock {C}omputational {T}hinking {T}est for {B}eginners: {D}esign and
  {C}ontent {V}alidation.
\newblock In {\em {EDUCON}}, 2020.

\bibitem{gonzalez2015computational}
Marcos~Rom{\'a}n Gonz{\'a}lez.
\newblock {C}omputational {T}hinking {T}est: {D}esign {G}uidelines and
  {C}ontent {V}alidation.
\newblock In {\em EDULEARN 2015}, 2015.

\bibitem{DBLP:journals/ce/TangYLHZ20}
Xiaodan Tang, Yue Yin, Qiao Lin, Roxana Hadad, and Xiaoming Zhai.
\newblock {A}ssessing {C}omputational {T}hinking: {A} {S}ystematic {R}eview of
  {E}mpirical {S}tudies.
\newblock {\em Computers \& Education}, 2020.

\bibitem{roman2019combining}
Marcos Rom{\'a}n-Gonz{\'a}lez, Jes{\'u}s Moreno-Le{\'o}n, and Gregorio Robles.
\newblock {C}ombining {A}ssessment {T}ools for a {C}omprehensive {E}valuation
  of {C}omputational {T}hinking {I}nterventions.
\newblock {\em Computational Thinking Education}, 2019.

\bibitem{DBLP:journals/chb/Roman-GonzalezP17}
Marcos Rom{\'{a}}n{-}Gonz{\'{a}}lez, Juan{-}Carlos
  P{\'{e}}rez{-}Gonz{\'{a}}lez, and Carmen Jim{\'{e}}nez{-}Fern{\'{a}}ndez.
\newblock {W}hich {C}ognitive {A}bilities {U}nderlie {C}omputational
  {T}hinking? {C}riterion {V}alidity of the {C}omputational {T}hinking {T}est.
\newblock {\em Computers in Human Behavior}, 72:678--691, 2017.

\bibitem{DBLP:journals/jeric/Lai22}
Rina P.~Y. Lai.
\newblock {B}eyond {P}rogramming: {A} {C}omputer-{B}ased {A}ssessment of
  {C}omputational {T}hinking {C}ompetency.
\newblock {\em Transactions on Computing Education}, 2022.

\bibitem{DBLP:conf/sigcse/BockmonB23}
Ryan Bockmon and Chris Bourke.
\newblock {V}alidation of the {P}lacement {S}kill {I}nventory: {A} {CS0/CS1}
  {P}lacement {E}xam.
\newblock In {\em {SIGCSE}}, 2023.

\bibitem{bloom2020taxonomy}
Benjamin~S Bloom and David~R Krathwohl.
\newblock {\em {T}axonomy of {E}ducational {O}bjectives: {T}he {C}lassification
  of {E}ducational {G}oals. {B}ook 1, {C}ognitive {D}omain}.
\newblock 2020.

\bibitem{hourofcode_maze}
Code.org.
\newblock {H}our of {C}ode: {C}lassic {M}aze {C}hallenge.
\newblock \url{https://studio.code.org/s/hourofcode}, 2022.

\bibitem{relkin2020techcheck}
Emily Relkin, Laura de~Ruiter, and Marina~Umaschi Bers.
\newblock {T}ech{C}heck: {D}evelopment and {V}alidation of an {U}nplugged
  {A}ssessment of {C}omputational {T}hinking in {E}arly {C}hildhood
  {E}ducation.
\newblock {\em Journal of Science Education and Technology}, 2020.

\bibitem{DBLP:journals/csedu/GaneIEYLP21}
Brian~D. Gane, Maya Israel, Noor Elagha, Wei Yan, Feiya Luo, and James~W.
  Pellegrino.
\newblock {D}esign and {V}alidation of {L}earning {T}rajectory-{B}ased
  {A}ssessments for {C}omputational {T}hinking in {U}pper {E}lementary
  {G}rades.
\newblock {\em Computer Science Education}, 2021.

\bibitem{DBLP:journals/ce/ChenSBJHE17}
Guanhua Chen, Ji~Shen, Lauren Barth{-}Cohen, Shiyan Jiang, Xiaoting Huang, and
  Moataz Eltoukhy.
\newblock {A}ssessing {E}lementary {S}tudents' {C}omputational {T}hinking in
  {E}veryday {R}easoning and {R}obotics {P}rogramming.
\newblock {\em Computers \& Education}, 2017.

\bibitem{DBLP:conf/sigcse/ParkerKSFKRW21}
Miranda~C. Parker, Yvonne~S. Kao, Dana Saito{-}Stehberger, Diana Franklin,
  Susan Krause, Debra~J. Richardson, and Mark Warschauer.
\newblock {D}evelopment and {P}reliminary {V}alidation of the {A}ssessment of
  {C}omputing for {E}lementary {S}tudents {(ACES)}.
\newblock In {\em {SIGCSE}}, 2021.

\bibitem{DBLP:conf/icer/WeintropW15}
David Weintrop and Uri Wilensky.
\newblock {U}sing {C}ommutative {A}ssessments to {C}ompare {C}onceptual
  {U}nderstanding in {B}locks-based and {T}ext-based {P}rograms.
\newblock In {\em {ICER}}, 2015.

\bibitem{DBLP:conf/wipsce/MuhlingRH15}
Andreas M{\"{u}}hling, Alexander Ruf, and Peter Hubwieser.
\newblock {D}esign and {F}irst {R}esults of a {P}sychometric {T}est for
  {M}easuring {B}asic {P}rogramming {A}bilities.
\newblock In {\em Workshop in {P}rimary and {S}econdary {C}omputing
  {E}ducation}, 2015.

\bibitem{pattis1981karel}
Richard~E Pattis.
\newblock {\em {K}arel the {R}obot: {A} {G}entle {I}ntroduction to the {A}rt of
  {P}rogramming}.
\newblock John Wiley \& Sons, Inc., 1981.

\bibitem{DBLP:conf/icer/ZhiCBP19}
Rui Zhi, Min Chi, Tiffany Barnes, and Thomas~W. Price.
\newblock {E}valuating the {E}ffectiveness of {P}arsons {P}roblems for
  {B}lock-based {P}rogramming.
\newblock In {\em {ICER}}, 2019.

\bibitem{codeorg}
Code.org.
\newblock {Code.org}.
\newblock \url{https://code.org/}, 2022.

\bibitem{rasch1993probabilistic}
Georg Rasch.
\newblock {\em {P}robabilistic {M}odels for {S}ome {I}ntelligence and
  {A}ttainment {T}ests}.
\newblock Education Resources Information Center (ERIC), 1993.

\bibitem{distefano2005using}
Christine DiStefano and Brian Hess.
\newblock {U}sing {C}onfirmatory {F}actor {A}nalysis for {C}onstruct
  {V}alidation: {A}n {E}mpirical {R}eview.
\newblock {\em Journal of Psychoeducational Assessment}, 2005.

\bibitem{tavakol2011making}
Mohsen Tavakol and Reg Dennick.
\newblock {M}aking {S}ense of {C}ronbach's {A}lpha.
\newblock {\em International Journal of Medical Education}, 2011.

\bibitem{DBLP:journals/tkl/GuggemosSR23}
Josef Guggemos, Sabine Seufert, and Marcos Rom{\'{a}}n{-}Gonz{\'{a}}lez.
\newblock {C}omputational {T}hinking {A}ssessment - {T}owards {M}ore {V}ivid
  {I}nterpretations.
\newblock {\em Technology, Knowledge and Learning}, 2023.

\bibitem{sedgwick2012pearson}
Philip Sedgwick.
\newblock {P}earson’s {C}orrelation {C}oefficient.
\newblock {\em {B}ritish {M}edical {J}ournal}, 2012.

\bibitem{DBLP:conf/icer/SwidanHS18}
Alaaeddin Swidan, Felienne Hermans, and Marileen Smit.
\newblock {P}rogramming {M}isconceptions for {S}chool {S}tudents.
\newblock In {\em {ICER}}, 2018.

\bibitem{welch1947generalization}
Bernard~L Welch.
\newblock The {G}eneralization of {S}tudent's {P}roblem {W}hen {S}everal
  {D}ifferent {P}opulation {V}arlances are {I}nvolved.
\newblock {\em Biometrika}, 34(1-2):28--35, 1947.

\end{thebibliography}
\onecolumn 

\appendix
In this section, we present the details of \ourtest{}. Appendix~\ref{sec.appendix.ace} contains the instructions for the test and $21$ test items from \ourtest{}.\footnote{The visual grid for Q17 in the appendix has been simplified in comparison to the visual grid used in the original Q17 question shown in Figure~\ref{fig4:test_questions.ds03}; see discussion in Section~\ref{sec.results.rq2}.} Appendix~\ref{sec.appendix.ace.solution} provides answers to the test items.

\section{\ourtestheader{} Instructions and Test Items}\label{sec.appendix.ace}
\vspace{5mm}


\begin{tikzpicture}
\draw[thick] (0,7.5) rectangle (18,20);

\node[align=center, font=\bfseries\fontsize{12}{14}\selectfont] at (9,19.5) {Instructions};

\node[text width=17cm, align=justify, font=\fontsize{10}{12}\selectfont] at (8.5,18.5) {
  \begin{itemize}
    \item The assessment has $21$ questions. Each question is a multi-choice question with 4 answer choices and only of one them is correct.
  \end{itemize}
};
\node[text width=17cm, align=justify, font=\fontsize{10}{12}\selectfont] at (8.5,17.5) {
  \begin{itemize}
    \item The questions are based on tasks from the Hour of Code: Maze Challenge which can be found on the following website: \url{https://studio.code.org/hoc/1}. 
  \end{itemize}
};
\node[text width=17cm, align=justify, font=\fontsize{10}{12}\selectfont] at (8.5,16.5) {
 \begin{itemize}
    \item Below we recap some important elements of a typical task grid:
  \end{itemize}
};
\node at (10, 13.5) {\includegraphics[width=0.45\textwidth]{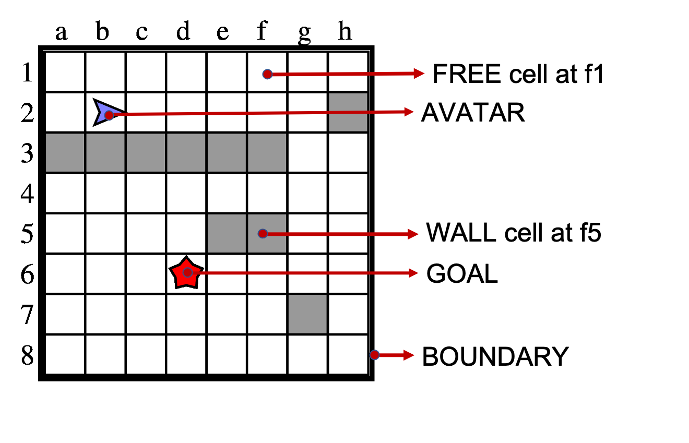}};
\node[text width=17cm, align=justify, font=\fontsize{10}{12}\selectfont] at (8.5,10.5) {
  \begin{itemize}
    \item AVATAR can face in the following four directions: \includegraphics[width=0.025\textwidth]{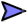}   (east),  \includegraphics[width=0.02\textwidth]{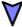} (south), \includegraphics[width=0.025\textwidth]{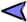} (west), \includegraphics[width=0.02\textwidth]{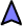} (north). AVATAR can move around the grid. Specifically, it moves in the direction of the arrow it is facing. It crashes if it tries to move into a WALL cell or BOUNDARY.       
      \end{itemize}
};
\node[text width=17cm, align=justify, font=\fontsize{10}{12}\selectfont] at (8.5,9) {
  \begin{itemize}
    \item A grid cell is denoted using the coordinates on the top (letters a--h) and left (numbers 1--8). For example, the AVATAR is at grid cell b2 and the GOAL is at grid cell d6. The grid cell at f5 is a WALL cell and the grid cell at f1 is a FREE cell. 
  \end{itemize}
};
\end{tikzpicture}

\clearpage
\begin{figure*}[t]
    \setlength{\fboxsep}{0.05pt}\fbox{\includegraphics[width=\textwidth]{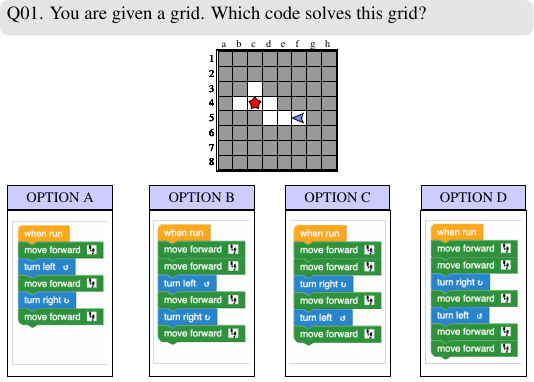}} 
\end{figure*}
\vspace*{\fill}
\clearpage

\begin{figure*}[t]
   \setlength{\fboxsep}{0.05pt}\fbox{\includegraphics[width=\textwidth]{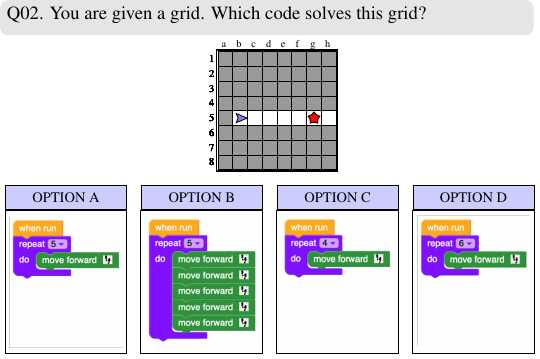}} 
\end{figure*}
\vspace*{\fill}
\clearpage

\begin{figure*}[t]
   \setlength{\fboxsep}{0.05pt}\fbox{\includegraphics[width=\textwidth]{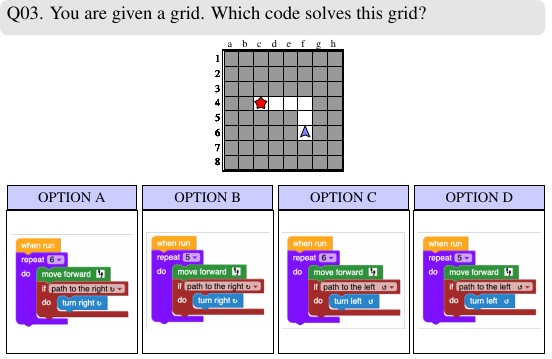}} 
\end{figure*}
\vspace*{\fill}
\clearpage

\begin{figure*}[t]
   \setlength{\fboxsep}{0.05pt}\fbox{\includegraphics[width=\textwidth]{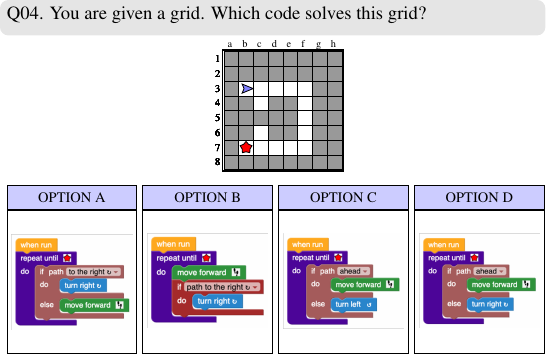}} 
\end{figure*}
\vspace*{\fill}
\clearpage

\begin{figure*}[t]
   \setlength{\fboxsep}{0.05pt}\fbox{\includegraphics[width=\textwidth]{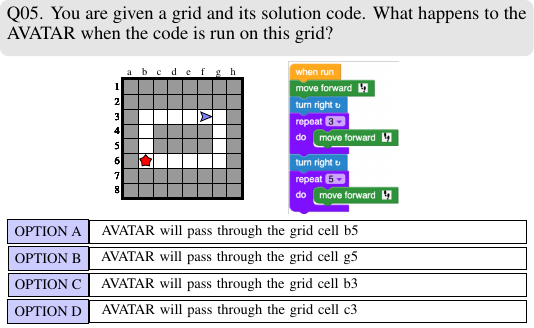}} 
\end{figure*}
\vspace*{\fill}
\clearpage

\begin{figure*}[t]
   \setlength{\fboxsep}{0.05pt}\fbox{\includegraphics[width=\textwidth]{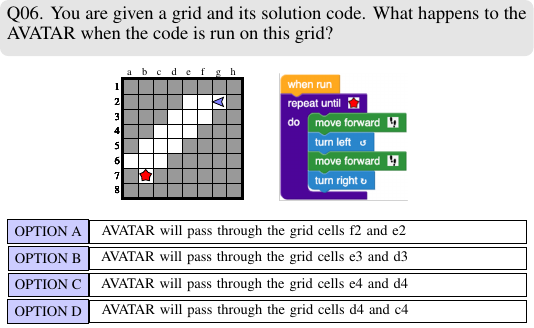}} 
\end{figure*}
\vspace*{\fill}
\clearpage

\begin{figure*}[t]
   \setlength{\fboxsep}{0.05pt}\fbox{\includegraphics[width=\textwidth]{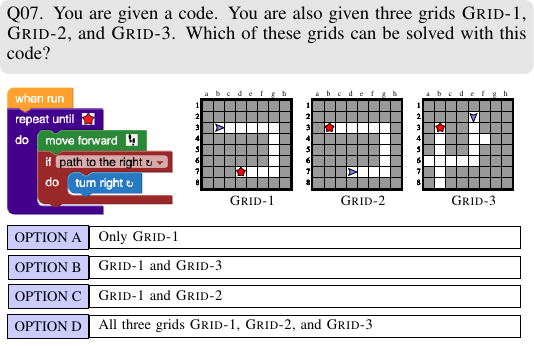}} 
\end{figure*}
\vspace*{\fill}
\clearpage

\begin{figure*}[t]
   \setlength{\fboxsep}{0.05pt}\fbox{\includegraphics[width=\textwidth]{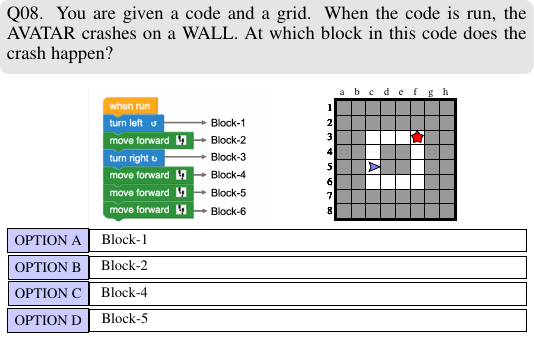}} 
\end{figure*}
\vspace*{\fill}
\clearpage

\begin{figure*}
   \setlength{\fboxsep}{0.05pt}\fbox{\includegraphics[width=\textwidth]{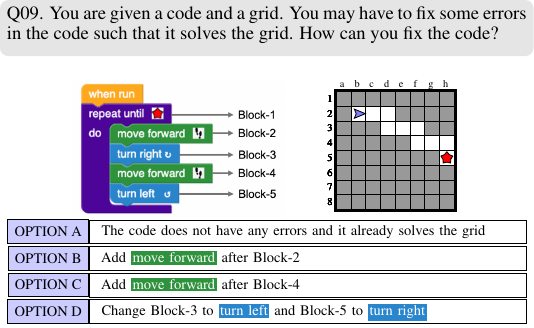}} 
\end{figure*}
\vspace*{\fill}
\clearpage

\begin{figure*}[t]
   \setlength{\fboxsep}{0.05pt}\fbox{\includegraphics[width=\textwidth]{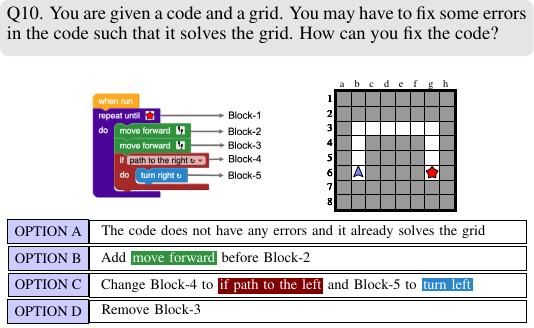}} 
\end{figure*}
\vspace*{\fill}
\clearpage

\begin{figure*}[t]
   \setlength{\fboxsep}{0.05pt}\fbox{\includegraphics[width=\textwidth]{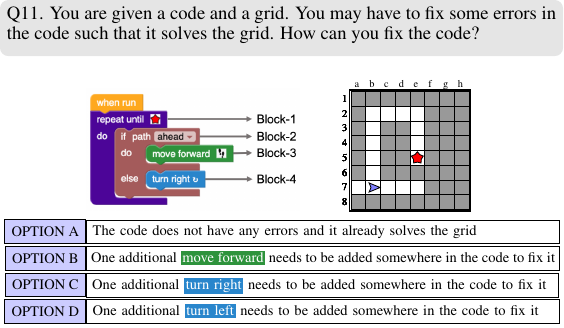}} 
\end{figure*}
\vspace*{\fill}
\clearpage

\begin{figure*}[t]
   \setlength{\fboxsep}{0.05pt}\fbox{\includegraphics[width=\textwidth]{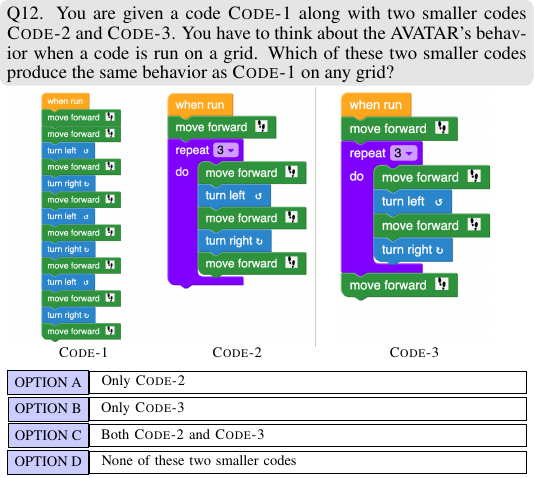}} 
\end{figure*}
\vspace*{\fill}
\clearpage

\begin{figure*}[t]
   \setlength{\fboxsep}{0.05pt}\fbox{\includegraphics[width=\textwidth]{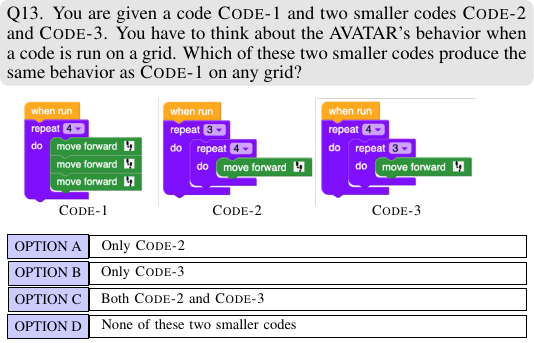}} 
\end{figure*}
\vspace*{\fill}
\clearpage

\begin{figure*}[t]
   \setlength{\fboxsep}{0.05pt}\fbox{\includegraphics[width=\textwidth]{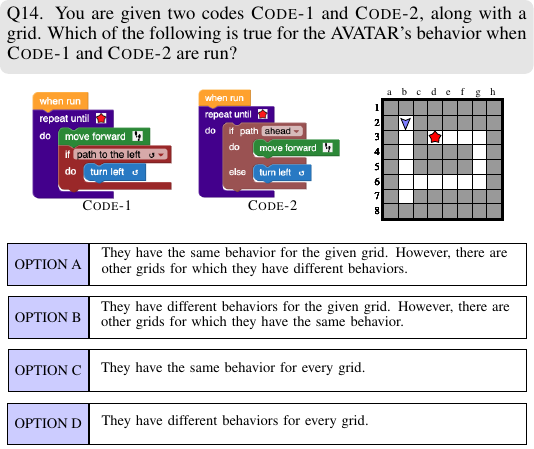}} 
\end{figure*}
\vspace*{\fill}
\clearpage

\begin{figure*}[t]
   \setlength{\fboxsep}{0.05pt}\fbox{\includegraphics[width=\textwidth]{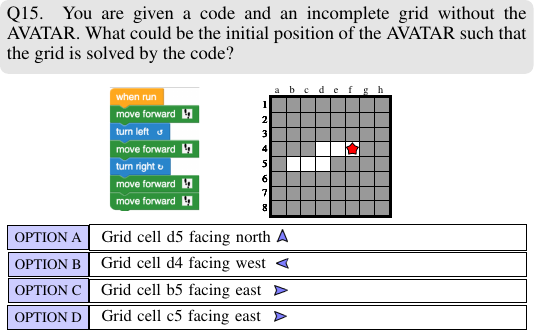}} 
\end{figure*}
\vspace*{\fill}
\clearpage

\begin{figure*}[t]
   \setlength{\fboxsep}{0.05pt}\fbox{\includegraphics[width=\textwidth]{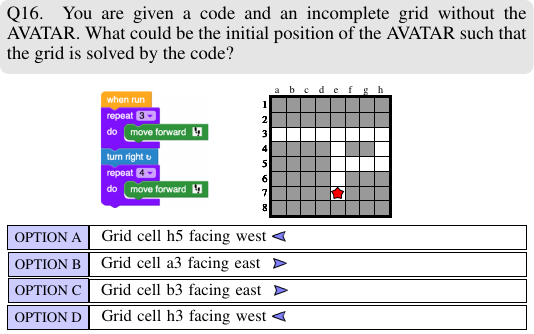}} 
\end{figure*}
\vspace*{\fill}
\clearpage

\begin{figure*}[t]
   \setlength{\fboxsep}{0.05pt}\fbox{\includegraphics[width=\textwidth]{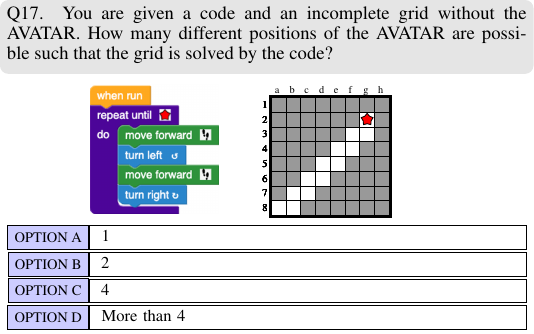}}
\end{figure*}
\vspace*{\fill}
\clearpage

\begin{figure*}[t]
   \setlength{\fboxsep}{0.05pt}\fbox{\includegraphics[width=\textwidth]{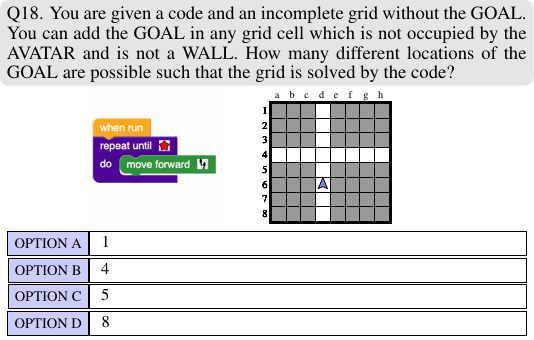}} 
\end{figure*}
\vspace*{\fill}
\clearpage

\begin{figure*}[t]
   \setlength{\fboxsep}{0.05pt}\fbox{\includegraphics[width=\textwidth]{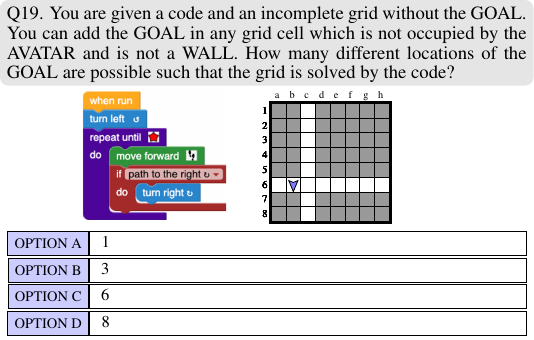}} 
\end{figure*}
\vspace*{\fill}
\clearpage

\begin{figure*}[t]
   \setlength{\fboxsep}{0.05pt}\fbox{\includegraphics[width=\textwidth]{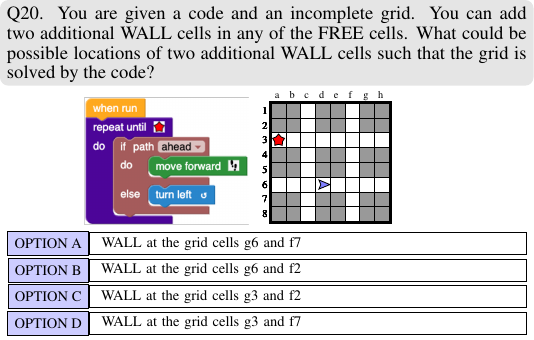}}
\end{figure*}
\vspace*{\fill}
\clearpage

\begin{figure*}[t]
   \setlength{\fboxsep}{0.05pt}\fbox{\includegraphics[width=\textwidth]{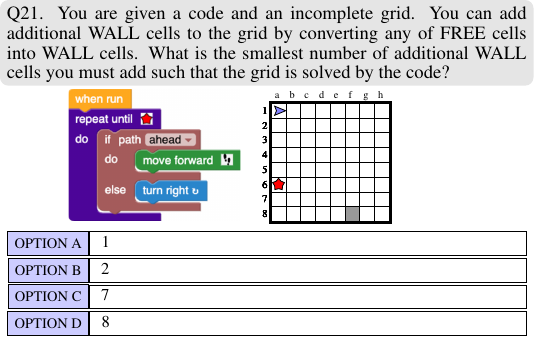}} 
\end{figure*}
\vspace*{\fill}
\clearpage

\section{Answers to \ourtestheader{} Test Items}\label{sec.appendix.ace.solution}

Below we provide answers to the $21$ \ourtest{} test items.

\begin{itemize}
    \item Q01: C
    \item Q02: A
    \item Q03: D
    \item Q04: D
    \item Q05: B
    \item Q06: C
    \item Q07: B
    \item Q08: C
    \item Q09: B
    \item Q10: D
    \item Q11: D
    \item Q12: B
    \item Q13: C
    \item Q14: B
    \item Q15: D
    \item Q16: C
    \item Q17: D
    \item Q18: C
    \item Q19: B
    \item Q20: B
    \item Q21: A
\end{itemize}



\end{document}